\documentclass[11pt]{article}%
\usepackage{amssymb}
\usepackage{amsfonts}
\usepackage{amsmath}
\usepackage{graphicx}%
\providecommand{\U}[1]{\protect\rule{.1in}{.1in}}
\newtheorem{theorem}{Theorem}

\begin{document}

\title{Universality of the SIS prevalence in networks}
\author{P. Van Mieghem\thanks{Delft University of Technology, Faculty of EECS, P.O Box
5031, 2600 GA Delft, The Netherlands; \emph{email}:
P.F.A.VanMieghem@tudelft.nl }}
\date{5 December 2016: for Nathan's birthday}
\maketitle

\begin{abstract}
Epidemic models are increasingly used in real-world networks to understand
diffusion phenomena (such as the spread of diseases, emotions, innovations,
failures) or the transport of information (such as news, memes in social
on-line networks). A new analysis of the prevalence, the expected number of
infected nodes in a network, is presented and physically interpreted. The
analysis method is based on spectral decomposition and leads to a universal,
analytic curve, that can bound the time-varying prevalence in any finite time
interval. Moreover, that universal curve also applies to various types of
Susceptible-Infected-Susceptible (SIS) (and Susceptible-Infected-Removed
(SIR)) infection processes, with both homogenous and heterogeneous infection
characteristics (curing and infection rates), in temporal and even
disconnected graphs and in SIS processes with and without self-infections. The
accuracy of the universal curve is comparable to that of well-established
mean-field approximations.

\end{abstract}

\section{Introduction to SIS epidemics on networks}

Epidemic processes on a network can approximately describe an amazingly large
variety of real-world processes \cite{PVM_RMP_epidemics2014}, such as the
spread of a disease, a digital virus, a message in an on-line social network,
an emotion, the propagation of a failure or an innovation and other diffusion
phenomena on networks (competing opinions, social contagion
\cite{Gleeson_PRX2016}). While the study of epidemics dates back to the great
Bernoulli's, the investigation of the role of the underlying graph on the
dynamics of the susceptible-infected-susceptible (SIS) process was only
initiated 15 years ago with the seminal paper of Pastor-Satorras and
Vespignani \cite{PastorSatorras_Vespignani_PRE2001}. The relatively new field
of network science
\cite{Watts,Dorogovtsev_Mendes2003book,Barrat_Bartelemy_Vespignani_CUPbook2008,Newman_boek2010,PVM_PAComplexNetsCUP,Barabasi_NetworkScience}
aims to study the interplay between dynamic processes on a graph and the
characteristics of that underlying graph. In \cite{PVM_accuracyNIMFA_2015}, we
discussed the \textquotedblleft Local rule-Global emergent
properties\textquotedblright\ (LrGep) class, where the collective action of
the local rules executed at each node in the network gives rise to a complex,
emergent global network behavior. Prominent examples of the LrGep-class are
epidemic models and more general reaction-diffusion processes
\cite{PVM_RMP_epidemics2014}, the Ising spin model \cite{Onsager1944}, the
Kuramoto coupled-oscillator model \cite{Strogatz_PhysicaD2000}, cellular
automata \cite{Wolfram_2002}, sandpiles as models for self-organized
criticality \cite{Bak_PRL1987,Goh_PRL2003,PVM_Tom_selforganization2009} and
opinion models \cite{Castellano_RMP2009,Li_opinion2013}. The fascinating
binding of these LrGep class members is that many LrGep models feature a phase
transition \cite{Stanley_phase_transition}, they all depend heavily on the
underlying topology and many processes in nature seem well described by
LrGep\ models. The simplicity of the local rules disguises the overwhelming
complexity of the global emergent network behavior that these local rules
create. Even one of the simplest members of the LrGep class, the SIS process,
is intricate and not sufficiently understood. However, the Markovian SIS
process on networks allows for the highest degree of analytic treatment, which
is a major motivation for the continued effort towards its satisfactory
understanding. Here, we report on a universal property of the SIS prevalence
and we propose a new analytic approximation with an accuracy comparable to the
well-established mean-field models \cite{PVM_RMP_epidemics2014}.

We consider an unweighted, undirected graph $G$ containing a set $\mathcal{N}$
of $N$ nodes and a set $\mathcal{L}$ of $L$ links. The topology of the graph
$G$ is represented by a symmetric $N\times N$ adjacency matrix $A$. In an SIS
epidemic process
\cite{Bailey_book,Anderson_May,Daley,Diekmann_Heesterbeek_Britton_boek2012,PVM_RMP_epidemics2014,Kiss_Miller_Simon2016}%
, the viral state of a node $i$ at time $t$ is specified by a Bernoulli random
variable $X_{i}\left(  t\right)  \in\{0,1\}$: $X_{i}\left(  t\right)  =0$,
when node $i$ is healthy, but susceptible and $X_{i}\left(  t\right)  =1$,
when node $i$ is infected. A node $i$ can only be in one of these two states:
\emph{infected}, with probability $\Pr[X_{i}(t)=1]$ or \emph{healthy}, with
probability $\Pr[X_{i}(t)=0]$, but susceptible to the infection. We assume
that the curing process for node $i$ is a Poisson process with rate
$\delta_{i}$ and that the infection rate over the link $\left(  i,j\right)  $
is a Poisson process with rate $\beta_{ij}$. Only when node $i$ is infected,
it can infect each node $k$ of its healthy direct neighbors with rate
$\beta_{ik}$. All Poisson curing and infection processes are independent. This
description defines the continuous-time, Markovian \emph{heterogeneous} SIS
epidemic process on a graph $G$. We do not consider non-Markovian epidemics
\cite{PVM_nonMarkovianSIS_2013,PVM_nonMarkovianSIS_NIMFA_2013} and assume that
the infection characteristics in the graph, i.e. all curing and infection
rates, are independent of time. The fraction of infected nodes is defined as%
\begin{equation}
S\left(  t\right)  =\frac{1}{N}\sum_{i=1}^{N}X_{i}\left(  t\right)
\label{def_S_fraction_infected_nodes}%
\end{equation}
and its expectation, called the \emph{prevalence} or the order parameter,
equals%
\begin{equation}
y\left(  t\right)  =E\left[  S\left(  t\right)  \right]  =\frac{1}{N}%
\sum_{i=1}^{N}\Pr\left[  X_{i}\left(  t\right)  =1\right]
\label{def_average_fraction_infected_nodes}%
\end{equation}
exploiting the property $E\left[  X_{i}\right]  =\Pr\left[  X_{i}=1\right]  $
of a Bernoulli distribution, which enables to avoid computations with the
probability operator in favor of the easier, linear expectation operator. In
that setting, the exact Markovian \emph{heterogeneous} SIS governing equation
\cite{PVM_secondorder_SISmeanfield_PRE2012,PVM_PAComplexNetsCUP} for the
infection probability of node $i$ is%
\begin{equation}
\frac{dE\left[  X_{i}\left(  t\right)  \right]  }{dt}=E\left[  -\delta
_{i}X_{i}\left(  t\right)  +\left(  1-X_{i}\left(  t\right)  \right)
\sum_{k=1}^{N}\beta_{ki}a_{ki}X_{k}\left(  t\right)  \right]
\label{governing_eq_heterogeneous_SIS}%
\end{equation}
When node $i$ is infected at time $t$ and $X_{i}\left(  t\right)  =1$, only
the first term on the right-hand side between the brackets $\left[  .\right]
$ affects and decreases with rate $-\delta_{i}$ the change in infection
probability with time $\frac{d\Pr\left[  X_{i}\left(  t\right)  =1\right]
}{dt}$ (left-hand side in (\ref{governing_eq_heterogeneous_SIS})). When node
$i$ is healthy, $X_{i}\left(  t\right)  =0$ and $\left(  1-X_{i}\left(
t\right)  \right)  =1$, only the second term between the brackets $\left[
.\right]  $ increases $\frac{d\Pr\left[  X_{i}\left(  t\right)  =1\right]
}{dt}$ by a rate $\sum_{k=1}^{N}\beta_{ki}a_{ki}X_{k}\left(  t\right)  $ due
to all its infected, direct neighbors. We define the Bernoulli random vector
$w\left(  t\right)  =\left(  X_{1}\left(  t\right)  ,X_{2}\left(  t\right)
,\ldots,X_{N}\left(  t\right)  \right)  $ at time $t$, the nodal curing vector
$\widetilde{\delta}=\left(  \delta_{1},\delta_{2},\ldots,\delta_{N}\right)  $
and the weighted adjacency matrix $\widetilde{A}$ with element $\widetilde
{a}_{ij}=\beta_{ij}a_{ij}$, that can change with time $t$ as in temporal
networks \cite{Holme_Saramki2012}. If $\beta_{ki}=\beta_{ik}$, the
corresponding \emph{heterogeneous} SIS prevalence differential equation is
(see Theorem \ref{theorem_governing_SIS_hetereogeneous_dvgl} in Appendix
\ref{sec_proof_basic_differential_equation})%
\begin{equation}
N\frac{dy\left(  t\right)  }{dt}=-E\left[  \widetilde{\delta}^{T}w\left(
t\right)  \right]  +E\left[  \left(  w\left(  t\right)  \right)
^{T}\widetilde{Q}\left(  t\right)  w\left(  t\right)  \right]
\label{dvgl_heterogeneous_prevalence}%
\end{equation}
where the time-depending, weighted Laplacian $\widetilde{Q}\left(  t\right)
=\widetilde{\Delta}\left(  t\right)  -\widetilde{A}\left(  t\right)  $ is an
$N\times N$ positive semi-definite symmetric matrix, with the diagonal matrix
$\widetilde{\Delta}\left(  t\right)  =$ diag$\left(  \widetilde{d}%
_{1},\widetilde{d}_{2},\ldots,\widetilde{d}_{N}\right)  $ and the infection
strength of node $k$ is $\widetilde{d}_{k}=\sum_{i=1}^{N}\beta_{ki}a_{ki}$. In
a \emph{homogeneous} SIS process, where all $\beta_{ij}=\beta$ and $\delta
_{j}=\delta$, (\ref{dvgl_heterogeneous_prevalence}) simplifies
\cite{PVM_bounds_SIS_prevalence} to,
\begin{equation}
\frac{dy\left(  t^{\ast};\tau\right)  }{dt^{\ast}}=-y\left(  t^{\ast}%
;\tau\right)  +\frac{\tau}{N}E\left[  w\left(  t^{\ast};\tau\right)
^{T}Qw\left(  t^{\ast};\tau\right)  \right]
\label{dvgl_av_numb_infections_SIS}%
\end{equation}
where $\tau=\frac{\beta}{\delta}$ is the effective infection rate, $t^{\ast
}=t\delta$ is the normalized time, $Q=\Delta-A$ is the Laplacian of the graph
$G$ with $\Delta=$ diag$\left(  d_{1},d_{2},\ldots,d_{N}\right)  $ and $d_{i}$
is the degree of node $i$. The corresponding governing equation for the
prevalence of the SIR process is deduced in
\cite{PVM_ExactMarkovianSIR_SIS_CDC2014} . Assume in a temporal network that
the infection characteristics do not change, but only links in the graph
change at time $t$: $A\left(  t-\varepsilon\right)  =A_{1}$ and $A\left(
t+\varepsilon\right)  =A_{2}$ for any arbitrarily small real $\varepsilon>0$.
Since the number $N$ of nodes does not change, the number of infected nodes is
continuous at time $t$. Thus, the Bernoulli vector $w\left(  t\right)
=\lim_{\varepsilon\rightarrow0}w\left(  t-\varepsilon\right)  =\lim
_{\varepsilon\rightarrow0}w\left(  t+\varepsilon\right)  $ is continuous at
time $t$ and the prevalence differential equation
(\ref{dvgl_heterogeneous_prevalence}) shows that%
\begin{equation}
\left.  \frac{dy\left(  t\right)  }{dt}\right\vert _{t+\varepsilon}-\left.
\frac{dy\left(  t\right)  }{dt}\right\vert _{t-\varepsilon}=\frac{1}%
{N}E\left[  \left(  w\left(  t\right)  \right)  ^{T}\left(  \widetilde{Q_{2}%
}-\widetilde{Q_{1}}\right)  w\left(  t\right)  \right]
\label{discontinuous_prevalence_derivative}%
\end{equation}
implying that the derivative of the prevalence is likely not continuous at the
time when the topology changes. On the other hand, the derivative
$\frac{dy\left(  t\right)  }{dt}$ is continuous when the topology does not
change (nor the infection characteristics). Thus, the SIS prevalence on
temporal networks may show a discontinuous slope at time $t$, from which a
topology change at that time $t$ may be inferred.%

\begin{figure}
[h]
\begin{center}
\includegraphics[
height=6.4185cm,
width=12.1166cm
]%
{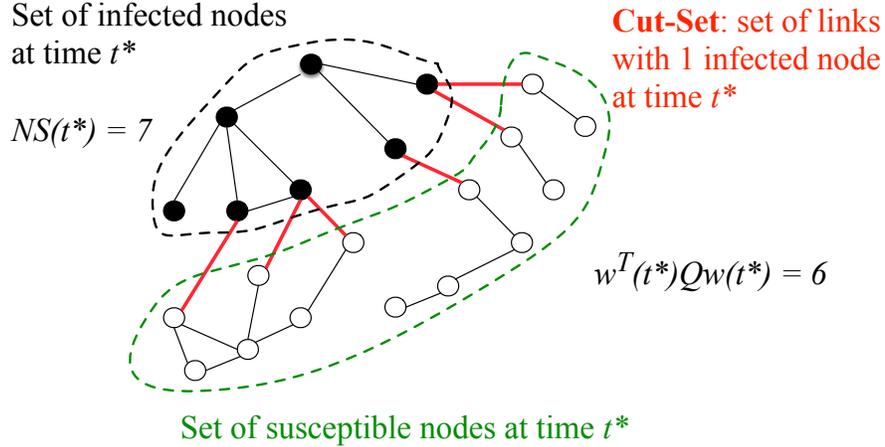}%
\caption{A sketch of an epidemic state in a graph at time $t^{\ast}$,
described by (\ref{dvgl_av_numb_infections_SIS}), illustrates the three sets:
(a) the set of infected nodes containing $NS\left(  t^{\ast}\right)  =7$ nodes
(in black), (b) the set of susceptible nodes (in green) and (c) the cut-set
(in red): the number of links with one infected node and here equal to
$w^{T}\left(  t^{\ast}\right)  Qw\left(  t\right)  =6$ links.}%
\label{Fig_sketch_cut_set}%
\end{center}
\end{figure}

\section{The cut-set}

The evolution of the nodal infection, described by
(\ref{governing_eq_heterogeneous_SIS}), reflects the \textquotedblleft local
rule\textquotedblright\ of the SIS process, whereas the SIS\ prevalence
differential equation (\ref{dvgl_heterogeneous_prevalence}) describes the
\textquotedblleft global emergent properties\textquotedblright. In
(\ref{governing_eq_heterogeneous_SIS}),\ (\ref{dvgl_heterogeneous_prevalence})
and (\ref{dvgl_av_numb_infections_SIS}), the second, non-linear term
quantifies the coupling between process and underlying topology.
Fig.~\ref{Fig_sketch_cut_set} illustrates that this physical interaction is
embedded in the cut-set $\left(  w\left(  t\right)  \right)  ^{T}Qw\left(
t\right)  $, which equals the number of links with one end node infected at
time $t$. Given that one node is infected initially and that the effective
infection rate $\tau$ is well above the epidemic threshold $\tau_{c}$, the
early infection spreads as the ripples in a pool caused by throwing a stone in
the water. First the direct neighbors become infected, then the neighbors of
those neighbors and so on. This early spread can be specified by the
\emph{expansion} of the graph \cite[p. 371]{PVM_PAComplexNetsCUP}, a graph
metric which determines the number of nodes at $k$ hops from the initial node.
In this early phase, the epidemic grows exponentially with time and the
cut-set boundary is analogous to \textquotedblleft concentric circles in a
pool\textquotedblright\ around the initially infected node
\cite{Brockmann_Science2013}. After some time, infected nodes cure and move in
Fig. \ref{Fig_sketch_cut_set} to the set of susceptible nodes. When the number
of successive shells around the initial node exceeds the average hopcount
\cite[p. 360 \& Chapter 16]{PVM_PAComplexNetsCUP}, i.e. number of links of the
shortest path between two arbitrary nodes, the finite size of the graph
prevents exponential increase in the number of nodes reached from an initial
node. Hence, two effects, curing and finite graph structure, limit the growth
of an epidemic. After the early phase, the cut-set as well as its border line
between infected and healthy nodes cease to resemble simple geometric
concentric circles and start exhibiting a complicated shape. Determining the
largest cut-set, which corresponds to the fastest possible viral increase (see
(\ref{dvgl_heterogeneous_prevalence})) in the network, is NP-hard, as well as
finding the smallest cut-size that is related \cite[p. 95]{PVM_graphspectra}
to the isoperimetric constant $\eta$, which upper bounds the epidemic
threshold $\tau_{c}\leq\frac{1}{\eta}$, as shown by Ganesh \emph{et al.
}\cite{Ganesh_2005}. In spite of its computational difficulty, the key to
understanding an infectious spread lies in the cut-set, which is the place to
prevent epidemic spread. The latest dynamic control strategies
\cite{Drakopoulos_CDC2015} target the reduction of the cut-set $\left(
w\left(  t\right)  \right)  ^{T}Qw\left(  t\right)  $ at each time $t$.

\section{Universality of the tanh-formula}

Our major new result concerns \textquotedblleft universality\textquotedblright%
: the time-varying prevalence $y\left(  t\right)  $ of any Markovian SIS
process, be it homogeneous or heterogeneous in its infection or/and curing
rates, in temporal or even disconnected graphs, with or without
self-infections, can be upper and lower bounded by a single, universal curve.
To simplify the explanation, we concentrate on a homogeneous SIS process and
refer to Appendices \ref{sec_tanh_heterogeneousSIS_k_components} and
\ref{sec_tanh_epsilonSIS_k_components} for the other cases.

Our method, which is \emph{entirely different from the mean-field concept}, is
based on the spectral decomposition of the cut-set $\left(  w\left(  t\right)
\right)  ^{T}Qw\left(  t\right)  $ and of the Bernoulli state vector $w$,
whose components $w_{j}=X_{j}$ are only zero or one. Physically, the dynamics
(\ref{dvgl_av_numb_infections_SIS}) of the SIS epidemics, characterized by the
Bernoulli vector $w$ and the Laplacian matrix $Q$, is mapped onto the
Laplacian eigenspace, determined by the underlying graph $G$. As shown in
Appendix \ref{sec_quadratic_form_z^TQz}, the Bernoulli vector $w$ is projected
onto the $N$ orthogonal axes formed by the real, normalized Laplacian
eigenvectors $x_{1},x_{2},\ldots,x_{N}$ belonging to the eigenvalues $\mu
_{1}\geq\mu_{2}\geq\cdots\geq\mu_{N}=0$, respectively, and obeying the
orthogonality condition $x_{k}^{T}x_{m}=1$ if $k=m$, otherwise $x_{k}^{T}%
x_{m}=0$. The coordinates in the Laplacian eigenvector basis, $\zeta_{j}%
=w^{T}x_{j}$ for $1\leq j\leq N$, completely specify the Bernoulli vector $w$.
The relative success of the method is, in contrast to the adjacency matrix,
due to the knowledge of one eigenvector $x_{N}=\frac{1}{\sqrt{N}}u$ belonging
to the eigenvalue $\mu_{N}=0$ and where $u=\left(  1,1,\ldots,1\right)  $ is
the all-one vector. The corresponding coordinate is $\zeta_{N}=\frac{1}%
{\sqrt{N}}u^{T}w=\sqrt{N}S$, by the definition
(\ref{def_average_fraction_infected_nodes}) written in vector form as
$S=\frac{u^{T}w}{N}$. Since $w$ is a zero-one vector, the largest scalar
product $\zeta_{j}=w^{T}x_{j}$ is $\zeta_{N}$, which means that the Bernoulli
vector $w$ is most close to the eigenvector $x_{N}$. In addition, the norm of
the Bernoulli vector equals the sum of its components, $w^{T}w=u^{T}w=NS$,
which allows to specify the second most influential coordinate $\zeta_{N-1}$.

We consider a graph $G$ consisting of $k$ connected components, where the $k$
smallest Laplacian eigenvalues are zero, but $\mu_{N-k}>0$ (see Appendix
\ref{sec_kernel_Q}). For a connected graph ($k=1$), the second smallest
eigenvalue $\mu_{N-1}>0$ of the Laplacian $Q$, coined by Fiedler
\cite{Fiedler1973} the algebraic connectivity, is studied over the last
decades \cite{PVM_graphspectra}. After spectral decomposition, the
differential equation (\ref{dvgl_av_numb_infections_SIS}) becomes%
\begin{equation}
\frac{dy\left(  t^{\ast};\tau\right)  }{dt^{\ast}}=\left(  \tau\mu
_{N-k}-1\right)  y\left(  t^{\ast};\tau\right)  -\tau\mu_{N-k}y^{2}\left(
t^{\ast};\tau\right)  -\Psi_{k}\left(  t^{\ast};\tau\right)
\label{prevalence_homogeneous_Riccati_like}%
\end{equation}
where the remainder $\Psi_{k}\left(  t^{\ast};\tau\right)  $ is explicitly
given in (\ref{explicit_remainder_Psi_k}). If $\Psi_{k}\left(  t^{\ast}%
;\tau\right)  $ equals a constant $c$, then
(\ref{prevalence_homogeneous_Riccati_like}) reduces to a Riccati differential
equation, which can be solved exactly. Assuming that we can bound $\Psi
_{k}\left(  t^{\ast};\tau\right)  $ in a normalized time interval $\left[
t_{1}^{\ast},t_{2}^{\ast}\right]  $ by two constants, $c_{L}\left(  k\right)
\leq$ $\Psi_{k}\left(  t^{\ast};\tau\right)  \leq c_{U}\left(  k\right)  $,
then the prevalence $y\left(  t^{\ast}\right)  $ can be bounded in $\left[
t_{1}^{\ast},t_{2}^{\ast}\right]  $, for the same initial condition $y_{0}$,
by
\[
T\left(  \left.  t^{\ast}\right\vert y_{0},\tau\mu_{N-k},c_{U}\left(
k\right)  \right)  \leq y\left(  t^{\ast}\right)  \leq T\left(  \left.
t^{\ast}\right\vert y_{0},\tau\mu_{N-k},c_{L}\left(  k\right)  \right)
\]
where our \textquotedblleft tanh-formula\textquotedblright\ is
\begin{equation}
T\left(  \left.  t\right\vert y_{0},s,c\right)  =\frac{1}{2}\left(  1-\frac
{1}{s}\right)  +\frac{\Xi}{2}\tanh\!\left(  \!\frac{s\Xi}{2}t+\Omega
_{0}\!\right)
\label{Riccati_approximation_prevalence_k_connected_component_graph}%
\end{equation}
with the Laplacian normalized effective infection rate $s=\tau\mu_{N-k}$ and%
\begin{equation}
\Omega_{0}=\text{arctanh}\!\left(  \frac{2y_{0}\!-\!\left(  \!1\!-\!\frac
{1}{s}\right)  }{\Xi}\!\right)  \label{def_Omega0}%
\end{equation}
and
\begin{equation}
\Xi=\sqrt{\left(  1-\frac{1}{s}\right)  ^{2}-\frac{4c}{s}}\label{def_Xi}%
\end{equation}

\begin{figure}
[h]
\begin{center}
\includegraphics[
height=12.1166cm,
width=8.6942cm
]%
{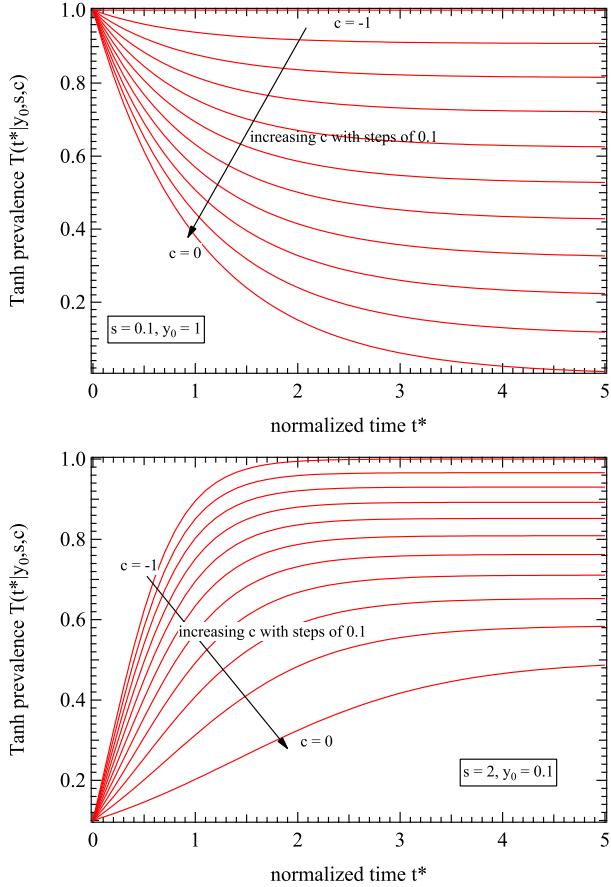}%
\caption{The tanh-formula
(\ref{Riccati_approximation_prevalence_k_connected_component_graph}) as a
function of normalized time $t^{\ast}$ for various values of $c=\left\{
-1,-0.9,-0.8,\ldots,0\right\}  $. Two characteristic regimes are shown:
(above) $s=\tau\mu_{N-k}$ below the epidemic threshold and initial condition
$y_{0}=1$, where all are nodes infected, and (below) $s=\tau\mu_{N-k}$ above
the epidemic threshold and initial condition $y_{0}=0.1.$}%
\label{Fig_tanh_c}%
\end{center}
\end{figure}

Fig. \ref{Fig_tanh_c} draws the tanh-formula
(\ref{Riccati_approximation_prevalence_k_connected_component_graph}) as a
function of normalized time $t^{\ast}$ for various $c\in\left[  -1,0\right]
$, in two characteristic regimes above ($\tau\mu_{N-k}$ high) and below
($\tau\mu_{N-k}$ small) the epidemic threshold.%

We argue in Appendix \ref{sec_tanh_homogenousSIS_k_components} that, for
$k=1$, a rough estimate for $c\approx-y_{0}$. Our tanh-formula
(\ref{Riccati_approximation_prevalence_k_connected_component_graph})
approximates the total contribution $\Psi_{k}\left(  t^{\ast};\tau\right)  $,
containing the less influential Bernoulli vector coordinates in the Laplacian
eigenspace, by a constant $c$. Extensive simulations
\cite{PVM_tanh_formula_prevalence} on various graph types and infection
characteristics, compared with the $N$-Intertwined Mean-Field Approximation
(NIMFA, \cite{PVM_N_intertwined_Computing2011}), demonstrate that the
tanh-formula
(\ref{Riccati_approximation_prevalence_k_connected_component_graph}) has an
overall performance comparable to mean-field approximations.

\section{Potential of the tanh-formula
(\ref{Riccati_approximation_prevalence_k_connected_component_graph})}

We discuss the tanh-formula
(\ref{Riccati_approximation_prevalence_k_connected_component_graph}) further.
First, (\ref{Riccati_approximation_prevalence_k_connected_component_graph})
contains three parameters: the initial condition $y_{0}$, $s=\tau\mu_{N-k}$
and $c$ that all depend upon the underlying graph. Remarkably, the expectation
of a complicated dynamic process -- the prevalence is an expectation -- is
approximately characterized by only three parameters. Second, the tanh-formula
(\ref{Riccati_approximation_prevalence_k_connected_component_graph})
generalizes the classical Kermack and McKendrick expression of 1927 by
incorporating the graph. Assuming \textquotedblleft homogeneous
mixing\textquotedblright, equivalent to regarding the underlying network as a
complete graph $K_{N}$, Kermack and McKendrick \cite{Kermack_McKendrick1927}
demonstrated that the SIR prevalence is described by the time-derivative of a
simplified variant of our tanh-formula
(\ref{Riccati_approximation_prevalence_k_connected_component_graph}). The
correspondence with our tanh-formula
(\ref{Riccati_approximation_prevalence_k_connected_component_graph}) is not so
surprising: in the complete graph $K_{N}$ with algebraic connectivity
$\mu_{N-1}=N$, the complicated remainder reduces to its simplest possible
form: $\Psi_{1}\left(  t^{\ast};\tau\right)  =$ $\tau N$Var$\left[  S\left(
t^{\ast};\tau\right)  \right]  $. In $K_{N}$, the fraction of infected nodes
$S\left(  t^{\ast};\tau\right)  $ is close to a Gaussian random variable
\cite{PVM_MSIS_star_PRE2012} above the epidemic threshold $\tau_{c}$ and the
variance Var$\left[  S\left(  t^{\ast};\tau\right)  \right]  \approx\frac
{1}{N}$ is almost constant in the metastable regime. Generally, in
sufficiently large networks and above the epidemic threshold, the average
total infection \textquotedblleft force\textquotedblright\ is balanced in
equilibrium by the average total healing \textquotedblleft
force\textquotedblright\ and the individual infection state $X_{j}$ of node
$j$ is only weakly dependent on $X_{k}$ of node $k$. Under these conditions of
weakly dependence, the Central Limit Theorem \cite{PVM_PAComplexNetsCUP}
states that the fraction $S$ of infected nodes tends to a Gaussian with mean
$y=E\left[  S\right]  $ and standard deviation $\sigma=\sqrt{\text{Var}\left[
S\right]  }$. The tanh-formula
(\ref{Riccati_approximation_prevalence_k_connected_component_graph}) does not
include the eventual die-out of the SIS epidemic in any finite network. Once
the epidemic has reached the metastable state in which the two above mentioned
forces balance each other on average, small process fluctuations of a couple
of standard deviations $\sigma$ around the mean $y$ continuously occur, but
large fluctuations are rare. In the metastable, a die-out of the SIS process
can only be caused by a cascade of mainly curing events in succession, which
is a very rare event. Consequently, once the process has reached the
metastable state, the epidemic remains in the network for a very long time
\cite{Draief_Massoulie,PVM_Survival_time_PRE2014,PVM_decay_SIS2014}, which
practically means for large real-world networks that the SIS epidemics remains
in the metastable state. Hence, for large $N$ and for effective infection
rates $\tau>\tau_{c}$, the tanh-formula
(\ref{Riccati_approximation_prevalence_k_connected_component_graph}) models
the \textquotedblleft real\textquotedblright\ SIS epidemic very well, although
it ignores absorption.
\begin{figure}
[h]
\begin{center}
\includegraphics[
height=7.2774cm,
width=12.1166cm
]%
{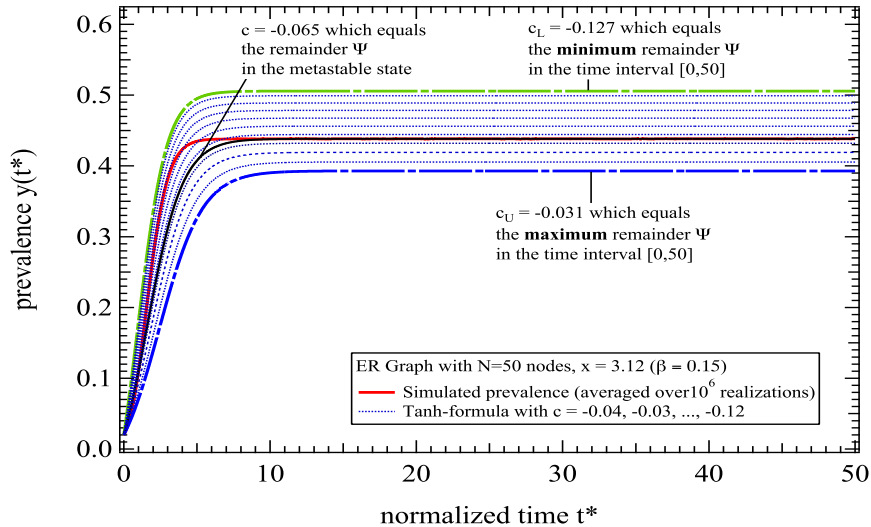}%
\caption{The prevalence envelope in a normalized time interval $\left[
0,50\right]  $ for an instance of an Erd\"{o}s-R\'{e}nyi random graph
$G_{p}\left(  N\right)  $ with $N=50$ nodes, link density $p=0.4$ and spectral
radius $\lambda_{1}=20.8$. Initially, one random node was infected. The
infection rate $\beta=0.15$ and the curing rate $\delta=1$, leading to a
normalized effective infection rate $x=\frac{\tau}{\tau_{c}^{\left(  1\right)
}}=\lambda_{1}\tau=3.12$.}%
\label{Fig_prevalence_c_sensitivity}%
\end{center}
\end{figure}
Third, the parameter $c$ approximates the complicated remainder $\Psi
_{k}\left(  t^{\ast};\tau\right)  $. The important bounding assumption
$c_{L}\left(  k\right)  \leq$ $\Psi_{k}\left(  t^{\ast};\tau\right)  \leq
c_{U}\left(  k\right)  $ leads to the prevalence envelope, illustrated in Fig.
\ref{Fig_prevalence_c_sensitivity} and akin to
\cite{PVM_topologicalRobustness_evalutation}, which encloses (see also Fig.
\ref{Fig_sisprevalence_realizations}) roughly 68\% of all realizations (i.e.
all possible real-world measurements of an SIS epidemic) assuming Gaussian
fluctuations around the prevalence -- which is, as mentioned above, a good
approximation for dense graphs as $K_{N}$ sufficiently above the epidemic
threshold. In absence of sufficiently clean data of a real-word SIS
prevalence, Fig. \ref{Fig_sisprevalence_realizations} plots 50 random
realizations of $S\left(  t^{\ast}\right)  $ out of $10^{6}$ with the same
infection characteristics and on the same graph as in Fig.
\ref{Fig_prevalence_c_sensitivity}. Since only the realizations that have
reached the metastable state after a start with one initially infected,
randomly chosen node, are observable, the prevalence is rescaled to
$y=\frac{N_{m}y_{m}+N_{d}y_{d}}{N}=\frac{N_{m}}{N}y_{m}$, where the index $m$
refers to those realizations that reach the metastable state, and $d$ those
that die out fast \cite{PVM_Die_out_Qiang2016} and never reach the metastable
state. Usually, the die-out probability, given an initial number of infected
nodes, is unknown, which complicates, as demonstrated in Fig.
\ref{Fig_sisprevalence_realizations} the proper normalization in reality,
where often only one realization of a spreading process (e.g. of a disease) is
measured over time. Fortunately, NIMFA \cite{PVM_bounds_SIS_prevalence} upper
bounds the prevalence, ignoring that realizations die out, while the
tanh-formula
(\ref{Riccati_approximation_prevalence_k_connected_component_graph}) can fit,
upper or lower bound data to infer from the parameters $\left(  y_{0},\tau
\mu_{N-k},c\right)  $ insights in the epidemic.
\begin{figure}
[h]
\begin{center}
\includegraphics[
height=7.2774cm,
width=12.1166cm
]%
{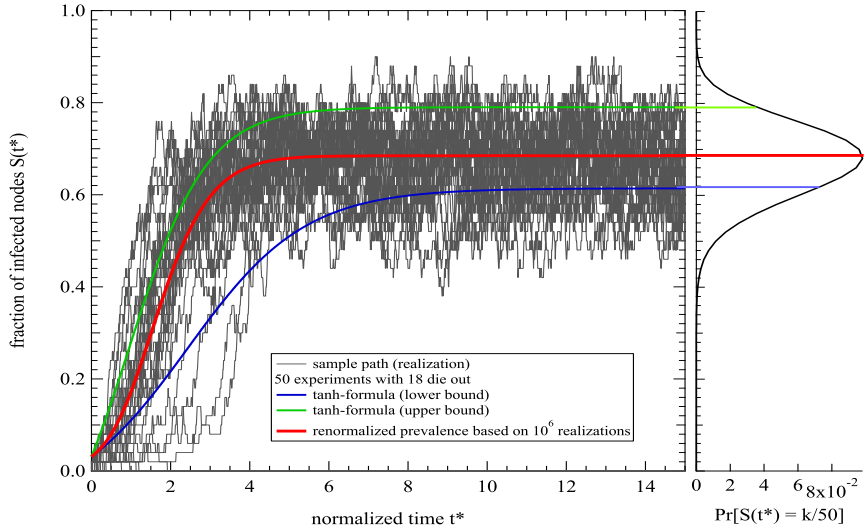}%
\caption{A random selection of 50 realizations out 10$^{6}$ realizations,
shown in Fig. \ref{Fig_prevalence_c_sensitivity}. Conditioned on 28
realization that did not die out, the prevalence (red) has been renormalized
as well as the tanh-formula
(\ref{Riccati_approximation_prevalence_k_connected_component_graph}) upper
(green) and lower bound (blue) in Fig. \ref{Fig_prevalence_c_sensitivity}}%
\label{Fig_sisprevalence_realizations}%
\end{center}
\end{figure}

Fourth, the tanh-formula
(\ref{Riccati_approximation_prevalence_k_connected_component_graph}) can be
used in temporal networks by gluing the different time-regimes in which the
network is unaltered: at time $t$, where the topology changes, we impose
continuity in the prevalence, $y(t-\varepsilon)=y\left(  t+\varepsilon\right)
$ for $\varepsilon\rightarrow0$, but allow discontinuity in the derivatives as
in (\ref{discontinuous_prevalence_derivative}). The tanh-formula
(\ref{Riccati_approximation_prevalence_k_connected_component_graph}) is the
more accurate, the better $\Psi_{k}\left(  t^{\ast};\tau\right)  $ can be
approximated by a constant. The smaller the time interval $\left[  t_{1}%
^{\ast},t_{2}^{\ast}\right]  $, the better $\Psi_{k}\left(  t^{\ast}%
;\tau\right)  $ is approximated by its mean $c=\frac{1}{t_{2}^{\ast}%
-t_{1}^{\ast}}\int_{t_{1}^{\ast}}^{t_{2}^{\ast}}\Psi_{k}\left(  t^{\ast}%
;\tau\right)  dt^{\ast}$. By dividing an experiment in small time intervals,
in which the data is fitted by the tanh-formula
(\ref{Riccati_approximation_prevalence_k_connected_component_graph}), and by
\textquotedblleft continuously gluing\textquotedblright\ the intervals that
determine $y_{0}$, a set $\left\{  \tau\mu_{N-k},c\right\}  $ for each
interval is obtained. Theory prescribes that all $\tau\mu_{N-k}$ over the
intervals should hardly differ, which can serve as an accuracy indication or a
verification that the epidemic process is Markovian SIS-like. The set of $c$
values then approximates the non-constant remainder $\Psi_{k}\left(  t^{\ast
};\tau\right)  $, that depends on both the epidemic process and the underlying graph.

Finally, just as higher order mean-field methods can increase the accuracy,
our spectral approach can be improved. Instead of bounding the remainder
$\Psi_{k}\left(  t^{\ast};\tau\right)  $ by a constant, which is the
zero-order approximation in the Taylor expansion of $\Psi_{k}\left(  t^{\ast
};\tau\right)  $, a polynomial in the prevalence $y$ seems promising
\cite{PVM_tanh_formula_prevalence}, which suggests that the method may be
refined further. Thus, we believe that it is worthwhile to research $\Psi
_{k}\left(  t^{\ast};\tau\right)  $ in depth to find sharper approximations.
Another extension towards more realistic \cite{PVM_Twitter_lognormal},
non-Poissonean infection and curing processes stands on the agenda for further research.

In summary, besides the powerful mean-field approximations, we have
demonstrated the potential of a spectral method for the prevalence to unravel
properties of SIS (SIR) epidemics on networks.

\medskip\textbf{Acknowledgement.} We are grateful to Qiang Liu for the
simulations that led to Fig. \ref{Fig_prevalence_c_sensitivity} and Fig.
\ref{Fig_sisprevalence_realizations}.

{\footnotesize
\bibliographystyle{unsrt}
\bibliography{cac,MATH,misc,net,pvm,QTH,tel}
}

\newpage

\appendix{}

\section{The basic differential equation of the SIS prevalence}

\label{sec_proof_basic_differential_equation}

\begin{theorem}
\label{theorem_governing_SIS_hetereogeneous_dvgl}Let $\widetilde{Q}\left(
t\right)  =\widetilde{\Delta}\left(  t\right)  -\widetilde{A}\left(  t\right)
$ denote the time-depending, weighted Laplacian, which is an $N\times N$
positive semi-definite symmetric matrix, with the diagonal matrix
$\widetilde{\Delta}\left(  t\right)  =$ diag$\left(  \widetilde{d}%
_{1},\widetilde{d}_{2},\ldots,\widetilde{d}_{N}\right)  $ and the infection
strength of node $k$ is $\widetilde{d}_{k}=\sum_{i=1}^{N}\beta_{ki}a_{ki}$. If
the link infection rate is the same in both directions, $\beta_{ki}=\beta
_{ik}$, then the corresponding \emph{heterogeneous} SIS prevalence
differential equation is%
\[
N\frac{dy\left(  t\right)  }{dt}=-E\left[  \widetilde{\delta}^{T}w\left(
t\right)  \right]  +E\left[  \left(  w\left(  t\right)  \right)
^{T}\widetilde{Q}\left(  t\right)  w\left(  t\right)  \right]
\]
If the link infection rate is not the same in both directions, $\beta_{ki}%
\neq\beta_{ik}$, a Laplacian representation is not possible and we end up with%
\[
N\frac{dy\left(  t\right)  }{dt}=-E\left[  \widetilde{\delta}^{T}w\left(
t\right)  \right]  +E\left[  \left(  u-w\left(  t\right)  \right)
^{T}\widetilde{A}w\left(  t\right)  \right]
\]
In a \emph{homogeneous} SIS epidemic process, where all $\beta_{ij}=\beta$ and
$\delta_{j}=\delta$, then (\ref{dvgl_heterogeneous_prevalence}) simplifies to
the differential equation,
\[
\frac{dy\left(  t^{\ast};\tau\right)  }{dt^{\ast}}=-y\left(  t^{\ast}%
;\tau\right)  +\frac{\tau}{N}E\left[  w\left(  t^{\ast};\tau\right)
^{T}Qw\left(  t^{\ast};\tau\right)  \right]
\]

\end{theorem}

\textbf{Proof:} Summing the Markovian \emph{heterogeneous} SIS governing
equation (\ref{governing_eq_heterogeneous_SIS}) for the infection probability
of node $i$ over all nodes and omitting the time-dependence in $X_{i}\left(
t\right)  $ to shorten the equations, yields%
\[
\frac{dE\left[  \sum_{i=1}^{N}X_{i}\right]  }{dt}=E\left[  -\sum_{i=1}%
^{N}\delta_{i}X_{i}+\sum_{k=1}^{N}\sum_{i=1}^{N}\beta_{ki}a_{ki}X_{k}%
-\sum_{k=1}^{N}\sum_{i=1}^{N}\beta_{ki}a_{ki}X_{i}X_{k}\right]
\]
After rewriting in matrix notation and in terms of the prevalence
(\ref{def_average_fraction_infected_nodes}), we obtain
\begin{align*}
N\frac{dy\left(  t;\tau\right)  }{dt}  &  =-E\left[  \widetilde{\delta}%
^{T}w\left(  t\right)  \right]  +E\left[  \left(  \widetilde{A}u\right)
^{T}w\left(  t\right)  -w\left(  t\right)  ^{T}\widetilde{A}w\left(  t\right)
\right] \\
&  =-E\left[  \widetilde{\delta}^{T}w\left(  t\right)  \right]  +E\left[
\left(  u-w\left(  t\right)  \right)  ^{T}\widetilde{A}w\left(  t\right)
\right]
\end{align*}

We define the weighted Laplacian as $\widetilde{Q}=\widetilde{\Delta
}-\widetilde{A}$, where the diagonal matrix $\widetilde{\Delta}=$ diag$\left(
\widetilde{d}_{1},\widetilde{d}_{2},\ldots,\widetilde{d}_{N}\right)  $ and the
strength of node $k$ is $d_{k}=\sum_{i=1}^{N}\beta_{ki}a_{ki}$. In order to
benefit from the basic Laplacian property $\widetilde{Q}u=0$ of a constant row
and column sum, we confine ourselves to a symmetric weighted adjacency matrix
$\widetilde{A}=\left(  \widetilde{A}\right)  ^{T}$, implying that $\beta
_{ij}=\beta_{ji}$. Thus, the infection rate of a link is only link dependent
and the same in both directions: from node $i$ to node $j$ and vice versa.
Consequently, the weighted Laplacian is symmetric, $\widetilde{Q}=\left(
\widetilde{Q}\right)  ^{T}$. Under this symmetry restriction, we have%
\begin{align*}
\left(  u-w\left(  t\right)  \right)  ^{T}\widetilde{A}w\left(  t\right)   &
=\left(  u-w\left(  t\right)  \right)  ^{T}\left(  \widetilde{\Delta
}-\widetilde{Q}\right)  w\left(  t\right) \\
&  =\left(  u-w\left(  t\right)  \right)  ^{T}\widetilde{\Delta}w\left(
t\right)  -u^{T}\widetilde{Q}w\left(  t\right)  +\left(  w\left(  t\right)
\right)  ^{T}\widetilde{Q}w\left(  t\right) \\
&  =\left(  w\left(  t\right)  \right)  ^{T}\widetilde{Q}w\left(  t\right)
\end{align*}
because%
\[
\left(  u-w\left(  t\right)  \right)  ^{T}\widetilde{\Delta}w\left(  t\right)
=\sum_{j=1}^{N}\left(  1-X_{j}\left(  t\right)  \right)  X_{j}\left(
t\right)  \widetilde{d}_{j}=0
\]
since $\left(  1-X_{j}\left(  t\right)  \right)  X_{j}\left(  t\right)  =0$ as
$X_{j}\left(  t\right)  \in\left\{  0,1\right\}  $. The homogeneous case
(\ref{dvgl_av_numb_infections_SIS}), as discussed in
\cite{PVM_bounds_SIS_prevalence}, follows directly from
(\ref{dvgl_heterogeneous_prevalence}) with normalized time $t^{\ast}=t\delta$.
\hfill$\square\medskip$

Perhaps surprising, the exact governing equations
(\ref{dvgl_heterogeneous_prevalence}) and (\ref{dvgl_av_numb_infections_SIS})
of the SIS prevalence are formally easier than their mean-field counterpart
(see \cite[p. 467]{PVM_PAComplexNetsCUP}).

By using the basic definition $Q=BB^{T}$ of the $N\times N$ Laplacian \cite[p.
14]{PVM_graphspectra} in terms of the $N\times L$ incidence matrix $B$, we
directly find \cite[p. 72]{PVM_graphspectra} that, for any $N\times1$ vector
$z$,%
\[
z^{T}Qz=\left(  B^{T}z\right)  ^{T}B^{T}z=\sum_{l\in\mathcal{L}}\left(
z_{l^{+}}-z_{l^{-}}\right)  ^{2}%
\]
where each link $l$ joins two end nodes $l^{+}$ and $l^{-}$. In particular,
for $z=w$, we observe that%
\begin{equation}
w^{T}Qw=\sum_{l\in\mathcal{L}}\left(  X_{l^{+}}-X_{l^{-}}\right)
^{2}\label{fundamental_Laplacian_quadratic_form}%
\end{equation}
If both end of a link $l$ are either infected or healthy, then $X_{l^{+}%
}-X_{l^{-}}=0$ and such a link $l$ does not contribute to the sum. Hence, only
links with one end infected and the other end healthy, for which $\left(
X_{l^{+}}-X_{l^{-}}\right)  ^{2}=1$, contribute precisely a unit amount to
$w^{T}Qw$. In other words, $w^{T}Qw$ equals the number of links in the cut-set.

\section{Kernel of the Laplacian $Q$ of a graph with $k$ disconnected
components}

\label{sec_kernel_Q}A graph $G$ has $k$ components (or clusters) if there
exists a relabeling of the nodes such that the adjacency matrix has the
structure%
\[
A=\left[
\begin{array}
[c]{cccc}%
A_{1} & O & \ldots & O\\
O & A_{2} &  & \vdots\\
\vdots &  & \ddots & \\
O &  & \ldots & A_{k}%
\end{array}
\right]
\]
where the square submatrix $A_{m}$ is the $n_{m}\times n_{m}$ adjacency matrix
of the connected component $m$ containing $n_{m}$ nodes. The total number of
nodes in $G$ equals $N=\sum_{m=1}^{k}n_{m}$. The corresponding Laplacian is%
\[
Q=\left[
\begin{array}
[c]{cccc}%
Q_{1} & O & \ldots & O\\
O & Q_{2} &  & \vdots\\
\vdots &  & \ddots & \\
O &  & \ldots & Q_{k}%
\end{array}
\right]
\]
Since $Q_{m}u_{n_{m}}=0$ for each connected component $m$ and the (unscaled)
$n_{m}\times1$ all-one eigenvector $u_{n_{m}}$ is the only eigenvector
belonging to eigenvalue $\mu_{n_{m}}=0$ (due to the connectivity of the
connected component $m$), we find that the general representation of an
eigenvector $x_{N}\left(  m\right)  $ of $Q$ belonging to the eigenvalue
$\mu_{N}=0$ with multiplicity $k$ is%
\begin{equation}
x_{N}^{T}\left(  m\right)  =\left(  \eta_{m1}u_{n_{1}}^{T},\eta_{m2}u_{n_{2}%
}^{T},\ldots,\eta_{mk}u_{n_{k}}^{T}\right)
\label{notation_m_degenerated_eigenvector}%
\end{equation}
The subspace of the $N$-dimensional space spanned by the eigenvectors of a
matrix belonging to the zero eigenvalue is called the kernel or null space of
that matrix. Each of the $k$ possible $N\times1$ eigenvectors $x_{N}\left(
m\right)  $ of $Q$ have the form (\ref{notation_m_degenerated_eigenvector})
and can thus be specified by a $k\times1$ vector $\eta_{m}=\left(  \eta
_{m1},\eta_{m2},\ldots,\eta_{mk}\right)  $ for $1\leq m\leq k$. Any pair
$\left\{  x_{N}\left(  m\right)  ,x_{N}\left(  s\right)  \right\}  $ of such
(normalized) eigenvectors, represented by the vectors $\eta_{m}=\left(
\eta_{m1},\eta_{m2},\ldots,\eta_{mk}\right)  $ and $\eta_{s}=\left(  \eta
_{s1},\eta_{s2},\ldots,\eta_{sk}\right)  $, must be orthogonal
\cite{PVM_graphspectra}, which leads to the set of $\binom{k}{2}$ non-linear
equations
\[
x_{N}\left(  m\right)  x_{N}^{T}\left(  s\right)  =\sum_{j=1}^{k}\eta_{mj}%
\eta_{sj}n_{j}=\delta_{ms}%
\]
Let us define the vector $y_{m}=\left(  y_{m1},y_{m2},\ldots,y_{mk}\right)  $
with $y_{mj}=\sqrt{n_{j}}\eta_{mj}$, then the above orthogonality condition
reduces to the \textquotedblleft ordinary\textquotedblright\ orthogonality
condition for the set of $k$ vectors $y_{1},y_{2},\ldots,y_{k}$,%
\[
y_{m}^{T}y_{s}=\sum_{j=1}^{k}y_{mj}y_{sj}=\delta_{ms}%
\]
This means that any set of $k$ orthogonal vectors $y_{1},y_{2},\ldots,y_{k}$,
that spans the $k$-dimensional space, can be used to produce $k$ eigenvectors
$x_{N}\left(  m\right)  $, with $1\leq m\leq k$, of the kernel of $Q$. The
corresponding $k\times k$ matrix $Y_{k}$, with the vectors $y_{1},y_{2}%
,\ldots,y_{k}$ in the columns, is an orthogonal matrix, whose properties with
respect to graphs are studied in \cite{PVM_eigenvectors_fundamentalweights}.
Clearly, the simplest set is the set $e_{1},e_{2},\ldots,e_{k}$ of the basis
vectors for which $Y_{k}=I$. Since any other $N\times1$ eigenvectors of $Q$
belonging to a positive eigenvalue of $Q$ is orthogonal to each kernel
eigenvector $x_{N}\left(  m\right)  $ that belongs to the zero eigenvalue
$\mu_{N}=0$, we observe that there exist a $k$-fold infinity of such
eigenvector sets, depending on our choice of the set of $k$ vectors
$y_{1},y_{2},\ldots,y_{k}$.

Since the Bernoulli vector $w$ has only zero and one components, the scalar
product $w^{T}x_{N}\left(  m\right)  $ is maximized if one the vectors $y_{m}$
is equal to the $k\times1$ all-one vector $u_{k}$. This observation suggests
us to construct the set of orthogonal eigenvectors $y_{1},y_{2},\ldots,y_{k}$,
with one of them, say $y_{1}=u_{k}$, equal to the all-one vector $u_{k}$.
Basically, this means that all orthogonal vectors $y_{1},y_{2},\ldots,y_{k}$
are eigenvectors of the adjacency matrix of the complete graph $K_{k}$,
because the unscaled largest adjacency eigenvector (of any regular graph and
thus also of $K_{k}$) is $x_{1}\left(  K_{k}\right)  =u_{k}$.

Barik \emph{et al.} \cite{Barik_LAA2011} have shown that only regular graphs,
such as the complete graph $K_{N}$, for $N=4k$ and $k\in\mathbb{N}_{0}$, and
the regular bipartite graph $K_{2k,2k}$, are diagonalizable by a symmetric
Hadamard matrix. An $n\times n$ Hadamard matrix $H_{n}$ contains as elements
either $-1$ and $1$ and obeys $H_{n}H_{n}^{T}=nI_{n}$. The normalized matrix
$\frac{1}{\sqrt{n}}H_{n}$ is an orthogonal matrix, from which it follows that
$\det H_{n}=n^{\frac{n}{2}}$, which is maximal among all $n\times n$ matrices
with elements in absolute value less than or equal to $1$, which includes all
orthogonal matrices. Indeed, let $H_{n}=\left[  u|\widetilde{H}\right]  $ so
that $H_{n}e_{1}=u_{n}$. Consider the diagonal matrix $D=I-e_{1}e_{1}^{T}$,
then%
\[
H_{n}DH_{n}^{T}=H_{n}H_{n}^{T}-H_{n}e_{1}\left(  H_{n}e_{1}\right)
^{T}=nI_{n}-u_{n}.u_{n}^{T}=nI-J
\]
Hence, the Laplacian matrix of the complete graph $K_{n}$ is $Q_{K_{n}%
}=nI-J=H_{n}DH_{n}^{T}$. Since $K_{n}$ is a regular graph, the eigenvectors of
the Laplacian $Q$ and the adjacency matrix $A$ are the same
\cite{PVM_graphspectra}. In conclusion, any Hadamard matrix with first column
$H_{n}e_{1}=u_{n}$ provides the orthogonal eigenvector matrix for the complete
graph $K_{n}$.

In summary, by choosing an $k\times k$ normalized Hadamard matrix $Y_{k}%
=\frac{1}{\sqrt{k}}H_{k}$ with first column $H_{k}e_{1}=u_{k}$, all components
$\eta_{mj}=\frac{y_{mj}}{\sqrt{n_{j}}}$ in
(\ref{notation_m_degenerated_eigenvector}) are determined, leading, with
$\left(  \Upsilon_{k}\right)  _{mj}=\eta_{mj}$, to the matrix $\Upsilon_{k}=$
diag$\left(  \frac{1}{\sqrt{n_{j}}}\right)  Y_{k}=\frac{1}{\sqrt{k}}%
$diag$\left(  \frac{1}{\sqrt{n_{j}}}\right)  H_{k}$. Moreover, the scalar
product $w^{T}\xi$ is maximal among all normalized vectors $\xi$ for
$\xi=x_{N}\left(  1\right)  $; in particular, $w^{T}x_{N}\left(  1\right)
=\frac{w^{T}u}{\sqrt{N}}=\frac{w^{T}w}{\sqrt{N}}$.

\section{The quadratic form $z^{T}Qz$ in a graph with $k$ disconnected
components and its spectral decomposition}

\label{sec_quadratic_form_z^TQz}

\subsection{The vector $z$ is real}

Since the eigenvectors of $Q$ constitute an orthogonal basis, any $N\times1$
real vector $z$ can be expressed as a linear combination of eigenvectors
$x_{1},x_{2},\ldots,x_{N}$ of $Q$,%
\[
z=\sum_{j=1}^{N}\alpha_{j}x_{j}%
\]
where $\alpha_{j}=z^{T}x_{j}$ and $x_{j}$ is the eigenvector belonging to the
$j$-th largest Laplacian eigenvalue $\mu_{j}$. In terms of the orthogonal
matrix $X$ with eigenvector in its columns
\cite{PVM_eigenvectors_fundamentalweights}, which satisfies the orthogonality
conditions $XX^{T}=X^{T}X=I$ so that $X^{-1}=X^{T}$, we have%
\[
\alpha=X^{T}z\text{ and }z=X\alpha
\]
illustrating the one-to-one relation between the coordinates of $z$ expressed
in a certain basis and its coordinates $\alpha$ expressed in the basis of
eigenvectors of $Q$. The quadratic form equals%
\begin{equation}
z^{T}Qz=\sum_{k=1}^{N}\sum_{m=1}^{N}\alpha_{k}\alpha_{m}x_{k}^{T}Qx_{m}%
=\sum_{k=1}^{N}\alpha_{k}^{2}\mu_{k} \label{quadratic_form_zTQz}%
\end{equation}

When the graph $G$ is disconnected into $k$ connected components (Appendix
\ref{sec_kernel_Q}), there holds \cite[p. 74]{PVM_graphspectra} that
$\mu_{N-j}=0$ for $0\leq j\leq k-1$. In other words, the $k$ smallest
eigenvalues of the Laplacian $Q$ are zero, whereas all the others are positive
$\mu_{1}\geq\mu_{2}\geq\ldots\geq$ $\mu_{N-k}>0$. Thus,%
\[
z^{T}z=\sum_{j=1}^{N}\alpha_{j}^{2}=\sum_{j=1}^{N-k}\alpha_{j}^{2}%
+\sum_{j=N-k+1}^{N}\alpha_{j}^{2}%
\]
Further, as shown in (\ref{notation_m_degenerated_eigenvector}) above for
$k>N-m$ and writing $z$ as a $k$-block vector,%
\[
z^{T}=\left(  \widehat{z}_{1}^{T},\widehat{z}_{2}^{T},\ldots,\widehat{z}%
_{k}^{T}\right)
\]
where $\widehat{z}_{j}$ is an $n_{j}\times1$ vector corresponding to the block
structure of the connected components in the adjacency matrix, the projection
of the vector $z$ onto the kernel vectors $x_{N}\left(  m\right)  $ for $1\leq
m\leq k$ of the Laplacian $Q$ is%
\[
z^{T}x_{N}\left(  m\right)  =\sum_{j=1}^{k}\eta_{mj}\left(  u_{n_{j}}%
^{T}\widehat{z}_{j}\right)
\]
so that%
\[
\sum_{j=N-k+1}^{N}\alpha_{j}^{2}=\sum_{m=1}^{k}\left(  \sum_{j=1}^{k}\eta
_{mj}\left(  u_{n_{j}}^{T}\widehat{z}_{j}\right)  \right)  ^{2}%
\]
In conclusion, for a graph $G$ with $k$ connected components, the Euclidean
norm of the vector $z$ can be written as
\begin{equation}
z^{T}z=\frac{1}{N}\left(  u^{T}z\right)  ^{2}+\sum_{m=2}^{k}\left(  \sum
_{j=1}^{k}\eta_{mj}\left(  u_{n_{j}}^{T}\widehat{z}_{j}\right)  \right)
^{2}+\sum_{j=1}^{N-k}\alpha_{j}^{2} \label{norm_k_connected_component_graph}%
\end{equation}
where each element $\eta_{mj}$ of the matrix $\Upsilon_{k}=\frac{1}{\sqrt{k}}%
$diag$\left(  \frac{1}{\sqrt{n_{j}}}\right)  H_{k}$ can be determined, as
shown in Appendix \ref{sec_kernel_Q}.

\subsection{The Bernoulli random vector $w$ is a binary vector}

The Bernoulli vector $w$ is a so-called binary vector, because each component
$w_{k}=X_{k}$ is either zero or one. For such vectors, we observe with
(\ref{def_S_fraction_infected_nodes}) that%
\[
w^{T}w=\sum_{k=1}^{N}X_{k}^{2}=\sum_{k=1}^{N}X_{k}=u^{T}w=NS
\]
Let us consider the eigenvector decomposition%
\begin{equation}
w=\sum_{j=1}^{N}\zeta_{j}x_{j} \label{w_as_lin_comb_xk}%
\end{equation}
where $\zeta_{j}=w^{T}x_{j}$ is the $j$-the coordinate of the Bernoulli vector
$w$ along the $j$-th eigenvector $x_{j}$ in the eigenspace of $Q$. For a graph
$G$ with $k$ connected components, (\ref{norm_k_connected_component_graph})
leads to
\[
w^{T}w=\frac{1}{N}\left(  u^{T}w\right)  ^{2}+\sum_{m=2}^{k}\left(  \sum
_{j=1}^{k}\eta_{mj}\left(  u_{n_{j}}^{T}w_{j}\right)  \right)  ^{2}+\sum
_{j=1}^{N-k}\zeta_{j}^{2}%
\]
or, with $w^{T}w=u^{T}w=NS$,%
\begin{equation}
\sum_{j=1}^{N-k}\zeta_{j}^{2}=N\left(  S-S^{2}\right)  -\sum_{m=2}^{k}\left(
\sum_{j=1}^{k}\eta_{mj}\left(  u_{n_{j}}^{T}w_{j}\right)  \right)  ^{2}
\label{sum_zeta^2_k_connected_component_graph}%
\end{equation}

Next, since $\mu_{N-j}=0$ for $0\leq j\leq k-1$, the quadratic form
(\ref{quadratic_form_zTQz}) becomes%
\begin{equation}
w^{T}Qw=\sum_{j=1}^{N-k}\zeta_{j}^{2}\mu_{j}
\label{quadratic_form_wTQw_k_connected_component_graph}%
\end{equation}
Introducing the square of the coordinate, obtained from
(\ref{sum_zeta^2_k_connected_component_graph}),%
\[
\zeta_{N-k}^{2}=N\left(  S-S^{2}\right)  -\sum_{m=2}^{k}\left(  \sum_{j=1}%
^{k}\eta_{mj}\left(  u_{n_{j}}^{T}w_{j}\right)  \right)  ^{2}-\sum
_{j=1}^{N-k-1}\zeta_{j}^{2}%
\]
into (\ref{quadratic_form_wTQw_k_connected_component_graph}) yields%
\begin{align*}
w^{T}Qw  &  =\zeta_{N-k}^{2}\mu_{N-k}+\sum_{j=1}^{N-k-1}\zeta_{j}^{2}\mu_{j}\\
&  =\mu_{N-k}N\left(  S-S^{2}\right)  -\mu_{N-k}\sum_{m=2}^{k}\left(
\sum_{j=1}^{k}\eta_{mj}\left(  u_{n_{j}}^{T}w_{j}\right)  \right)  ^{2}%
-\sum_{j=1}^{N-k-1}\mu_{N-k}\zeta_{j}^{2}+\sum_{j=1}^{N-k-1}\zeta_{j}^{2}%
\mu_{j}%
\end{align*}
Rewritten,%
\begin{equation}
w^{T}Qw=\mu_{N-k}N\left(  S-S^{2}\right)  +R_{k}
\label{quadratic_form_wTQw_inS_k_connected_component_graph}%
\end{equation}
where the correction $R_{k}$ is%
\begin{equation}
R_{k}=\sum_{j=1}^{N-k-1}\zeta_{j}^{2}\left(  \mu_{j}-\mu_{N-k}\right)
-\mu_{N-k}\sum_{m=2}^{k}\left(  \sum_{j=1}^{k}\eta_{mj}\left(  u_{n_{j}}%
^{T}\widehat{w}_{j}\right)  \right)  ^{2} \label{general_R_k_correction}%
\end{equation}
where $\widehat{w}_{j}$ here is the $j$-th $n_{j}\times1$ block vector of $w$
according to the component structure of the graph $G$. Thus, $u_{n_{j}}%
^{T}\widehat{w}_{j}$ equals the number of infected nodes at time $t$ in the
$j$-th connected component of $G$ with $n_{j}$ nodes and, with $N=\sum
_{m=1}^{k}n_{m}$, the fraction $S$ of infected nodes in $G$ (at time $t$) is a
\textquotedblleft weighted\textquotedblright\ average over the $k$ components
of $G$%
\[
S=\frac{\sum_{j=1}^{k}u_{n_{j}}^{T}\widehat{w}_{j}}{\sum_{m=1}^{k}n_{m}}%
\]

The first term in $R_{k}$ is non-negative, as well as the second term. In a
connected graph $G$ where $k=1$, the second term in
(\ref{general_R_k_correction}) vanishes so that $R_{1}$ is non-negative, but
only zero for the complete graph $K_{N}$. When $k=1$, the general expression
$R_{k}$ in (\ref{general_R_k_correction}) reduces to our previous expression
in \cite{PVM_bounds_SIS_prevalence} in terms of the algebraic connectivity
$\mu_{N-1}$. Clearly, the second term in (\ref{general_R_k_correction}) only
appears when a graph is disconnected into $k$ connected components. When $k$
is large, then $R_{k}$ is likely negative, and certainly for $k=N-1$, in which
case the first in (\ref{general_R_k_correction}) term vanishes.

For a given graph $G$, all parameters related its Laplacian eigenstructure,
such as the eigenvalues $\mu_{j}$ and the elements $\eta_{mj}$ of the Hadamard
related matrix $\Upsilon_{k}=\frac{1}{\sqrt{k}}$diag$\left(  \frac{1}%
{\sqrt{n_{j}}}\right)  H_{k}$, are known. Only the coordinates $\zeta_{j}$ for
$1\leq j\leq N-k$ and kernel coordinates $u_{n_{j}}^{T}\widehat{w}_{j}$ for
$1\leq j\leq k$ in (\ref{general_R_k_correction}) depend on the SIS process
via the Bernoulli vector $w\left(  t\right)  $, that depends upon its initial
value at $w\left(  0\right)  $ at time $t=0$. Indeed, if the initially
infected nodes only appear in one component, say $m=1$, so that the vector
$\widehat{w}_{1}\left(  0\right)  \neq0$, then all other vectors $\widehat
{w}_{m}=0$ for all components $m>1$, because evidently, an infection can only
spread in a connected component.

\subsection{A Fourier analysis interpretation of Laplacian eigenvectors}

If $x_{k}$ is an eigenvector of $Q$ belonging to eigenvector $\mu_{k}$, then
the fundamental Laplacian quadratic form in
(\ref{fundamental_Laplacian_quadratic_form}) becomes%
\[
\mu_{k}=x_{k}Qx_{k}=\sum_{l\in\mathcal{L}}\left(  \left(  x_{k}\right)
_{l^{+}}-\left(  x_{k}\right)  _{l^{-}}\right)  ^{2}%
\]
which implies that the variation of eigenvector components at both ends of a
link increases with the Laplacian eigenvalue. When interpreting the
eigenvector $x_{k}$ as a function $\left(  x_{k}\right)  _{i}$ of the nodal
components $i$ at frequency $\mu_{k}$, the above suggests that a high
frequency Laplacian eigenvector oscillates more (over a link) than a low
frequency Laplacian eigenvector. The suggestion is correct for a ring graph
\cite[p. 116-123]{PVM_graphspectra}, because the orthogonal eigenvector matrix
$X$ of the ring graph is the Fourier matrix (with the usual cosine and sine as
eigenfunctions). While a general theorem valid for any graph that the
Laplacian eigenvector $x_{k}$ possesses more sign changes with increasing
$\mu_{k}$ -- a reflection of higher oscillatory behavior with increasing
frequency --  seems missing\footnote{Any symmetric matrix can be reduced by
orthogonal Householder reflections to a tri-band matrix, whose eigenvector
structure consists of orthogonal polynomials and is computed in \cite[p.
565-573]{PVM_PAComplexNetsCUP},\cite{PVM_decay_SIS2014}.
\par
{}}, the intuition of the Fourier decomposition of a signal hints that the
\textquotedblleft Fourier coefficients\textquotedblright\ $\zeta_{k}%
=w^{T}x_{k}$ are generally expected to decrease with higher index $k$. If
correct and if $\zeta_{k}^{2}\mu_{k}$ decreases generally with $\mu_{k}$, then
the first sum of $R_{k}$ in (\ref{general_R_k_correction}) would generally
consists of decreasing positive terms. This interpretation may lead to sharp
approximations of $R_{k}$.

\section{Governing equation of the homogeneous SIS prevalence in graph with
$k$ disconnected components}

\label{sec_tanh_homogenousSIS_k_components}Invoking the definition
(\ref{def_average_fraction_infected_nodes}) of the prevalence $y=E\left[
S\right]  $, $E\left[  S-S^{2}\right]  =y-E\left[  S^{2}\right]  $ and
$E\left[  S^{2}\right]  =y^{2}+$ Var$\left[  S\right]  $,
(\ref{quadratic_form_wTQw_inS_k_connected_component_graph}) becomes%
\begin{align}
\frac{1}{N}E\left[  w^{T}Qw\right]   &  =\mu_{N-k}E\left[  S\right]
-\mu_{N-k}E\left[  S^{2}\right]  +\frac{E\left[  R_{k}\right]  }{N}\nonumber\\
&  =\mu_{N-k}y-\mu_{N-k}y^{2}-\mu_{N-k}\left(  \text{Var}\left[  S\right]
-\frac{E\left[  R_{k}\right]  }{N\mu_{N-k}}\right)
\label{E[w^t Q w]_homogeneous_k_disconnected_compon}%
\end{align}
Using (\ref{dvgl_av_numb_infections_SIS}), the spectral representation of the
SIS prevalence governing differential equation is%
\[
\frac{dy\left(  t^{\ast};\tau\right)  }{dt^{\ast}}=\left(  \tau\mu
_{N-k}-1\right)  y\left(  t^{\ast};\tau\right)  -\tau\mu_{N-k}y^{2}\left(
t^{\ast};\tau\right)  -\Psi_{k}\left(  t^{\ast};\tau\right)
\]
where the remainder%
\[
\Psi_{k}\left(  t^{\ast};\tau\right)  =\tau\mu_{N-k}\left(  \text{Var}\left[
S\left(  t^{\ast};\tau\right)  \right]  -\frac{E\left[  R_{k}\left(  t^{\ast
};\tau\right)  \right]  }{N\mu_{N-k}}\right)
\]
For a connected graph (i.e. with $k=1$ connected component), we again find the
earlier result in \cite[eq. (12)]{PVM_bounds_SIS_prevalence}. Introducing
(\ref{general_R_k_correction}), the explicit form of the remainder is%
\begin{equation}
\frac{\Psi_{k}\left(  t^{\ast};\tau\right)  }{\tau\mu_{N-k}}=\text{Var}\left[
S\left(  t^{\ast};\tau\right)  \right]  +\frac{1}{N}\sum_{m=2}^{k}E\left[
\left(  \sum_{j=1}^{k}\eta_{mj}\left(  u_{n_{j}}^{T}w_{j}\right)  \right)
^{2}\right]  -\frac{1}{N}\sum_{j=1}^{N-k-1}\left(  \frac{\mu_{j}}{\mu_{N-k}%
}-1\right)  E\left[  \zeta_{j}^{2}\right]  \label{explicit_remainder_Psi_k}%
\end{equation}
illustrating that $\Psi_{k}\left(  t^{\ast};\tau\right)  $ is likely positive
for large $k$, i.e. in a graph with many connected components.

For small time, the Taylor expansion yields%
\[
y\left(  \left.  t^{\ast}\right\vert \tau\right)  =y\left(  \left.
0\right\vert \tau\right)  +\left.  \frac{dy\left(  t^{\ast};\tau\right)
}{dt^{\ast}}\right\vert _{t^{\ast}=0}t^{\ast}+O\left(  t^{\ast2}\right)
\]
Invoking the differential equation (\ref{prevalence_homogeneous_Riccati_like})
and $y\left(  \left.  0\right\vert \tau\right)  =y_{0}$ leads to%
\[
y\left(  \left.  t^{\ast}\right\vert \tau\right)  =y_{0}+\left\{  y_{0}\left(
\tau\mu_{N\!-\!k}\!-\!1\right)  -y_{0}^{2}\!\tau\mu_{N\!-\!k}-\Psi_{k}\left(
0;\tau\right)  \right\}  t^{\ast}+O\left(  t^{\ast2}\right)
\]
where $\Psi_{1}\left(  0;\tau\right)  $ is small (because Var$\left[  S\left(
0;\tau\right)  \right]  =0$ since $S\left(  0\right)  $ is deterministic). The
tanh-approximation
(\ref{Riccati_approximation_prevalence_k_connected_component_graph}), on the
other hand, replaces $\Psi_{k}\left(  t^{\ast};\tau\right)  $ by $c$ (in any
time interval) so that
\[
T\left(  \left.  t^{\ast}\right\vert y_{0},\tau\mu_{N-k},c\right)
=y_{0}+y_{0}\left\{  \tau\mu_{N\!-\!k}\!\left(  1-y_{0}\right)  -\!\left(
1+\frac{c}{y_{0}}\right)  \right\}  t^{\ast}+O\left(  t^{\ast2}\right)
\]
The initial slope is non-negative when $\tau\geq\frac{1}{\mu_{N\!-\!k}%
}\!\left(  \frac{1+\frac{c}{y_{0}}}{1-y_{0}}\right)  $. For a connected graph
$\left(  k=1\right)  $, the tanh-prevalence $T\left(  \left.  t^{\ast
}\right\vert y_{0},\tau\mu_{N-1},c\right)  $ increases for $\tau>\tau_{c}$ for
all time $t^{\ast}\geq0$ and, hence,%
\[
\frac{1}{\mu_{N\!-\!1}}\!\left(  \frac{1+\frac{c}{y_{0}}}{1-y_{0}}\right)
>\tau_{c}%
\]
or%
\[
c>y_{0}\tau_{c}\mu_{N\!-\!1}\left(  1-y_{0}\right)  -y_{0}\geq y_{0}\frac
{\mu_{N\!-\!1}}{\lambda_{1}}\left(  1-y_{0}\right)  -y_{0}%
\]
because the lower bound for the epidemic threshold obeys $\tau_{c}\geq\frac
{1}{\lambda_{1}}$ (see e.g.
\cite{PVM_RMP_epidemics2014,PVM_bounds_SIS_prevalence}). Since $\mu
_{N-1}<\lambda_{1}$ (except for the complete graph and regular multipartite
graphs), the positive-slope condition would suggest that $c\gtrapprox-y_{0}$
for almost all graphs. The value $c\approx-y_{0}$ is also approximately
deduced from the Kermack and McKendrick \cite{Kermack_McKendrick1927} analysis
for SIR. Simulations \cite{PVM_tanh_formula_prevalence} seem to agree roughly
with $c\approx-y_{0}$.

For large time, the tanh-formula
(\ref{Riccati_approximation_prevalence_k_connected_component_graph}) reduces
to
\begin{equation}
\lim_{t^{\ast}\rightarrow\infty}T\left(  \left.  t^{\ast}\right\vert
y_{0},\tau\mu_{N-k},c\right)  =\frac{1}{2}\left(  1-\frac{1}{\tau\mu_{N-k}%
}\right)  +\frac{1}{2}\sqrt{\left(  1-\frac{1}{\tau\mu_{N-k}}\right)
^{2}-\frac{4c}{\tau\mu_{N-k}}} \label{tanh_steady_state}%
\end{equation}
which corresponds to the metastable state of the SIS process. For some
extremal values of $c$, (\ref{tanh_steady_state}) shows that%
\begin{align*}
\lim_{t^{\ast}\rightarrow\infty}T\left(  \left.  t^{\ast}\right\vert
y_{0},\tau\mu_{N-k},-1\right)   &  =1\\
\lim_{t^{\ast}\rightarrow\infty}T\left(  \left.  t^{\ast}\right\vert
y_{0},\tau\mu_{N-k},0\right)   &  =\left\{
\begin{array}
[c]{cc}%
1-\frac{1}{\tau\mu_{N-k}} & \tau>\frac{1}{\mu_{N-k}}\\
0 & \tau<\frac{1}{\mu_{N-k}}%
\end{array}
\right.
\end{align*}
For $c_{\operatorname{real}}=\frac{\tau\mu_{N-k}}{4}\left(  1-\frac{1}{\tau
\mu_{N-k}}\right)  ^{2}\geq0$, that guarantees a real prevalence ($\Xi=0$ in
(\ref{def_Xi})), the prevalence does not depend on time any more and $T\left(
\left.  t^{\ast}\right\vert y_{0},\tau\mu_{N-k},c_{\operatorname{real}%
}\right)  =\frac{1}{2}\left(  1-\frac{1}{\tau\mu_{N-k}}\right)  $. Moreover,
for $\tau<\frac{1}{\mu_{N-k}}$, positive $c>0$ are not physical since the
prevalence can become negative. For $\tau>\frac{1}{\mu_{N-k}}$, a positive
$c<c_{\operatorname{real}}$ can be possible.

Similarly to \cite{PVM_bounds_SIS_prevalence} for the case $k=1$ of a
connected graph, $T\left(  \left.  t^{\ast}\right\vert y_{0},s,c\right)  $ in
(\ref{Riccati_approximation_prevalence_k_connected_component_graph}) obeys the
Riccati differential equation%
\[
\frac{dT}{dt^{\ast}}=\left(  s-1\right)  T-sT^{2}-c
\]
For the same initial prevalence $y\left(  0\right)  =\tilde{y}_{k}\left(
0\right)  =y_{0}$ and given the constants $c_{L}\left(  k\right)  $ and
$c_{H}\left(  k\right)  $, that satisfy $c_{L}\left(  k\right)  \leq$
$\Psi_{k}\left(  t^{\ast};\tau\right)  \leq c_{U}\left(  k\right)  $, the
prevalence $y\left(  t^{\ast}\right)  $ at normalized time $t^{\ast}$ is
bounded by the relatively simple expression
(\ref{Riccati_approximation_prevalence_k_connected_component_graph})%
\[
\left\{
\begin{array}
[c]{cc}%
y\left(  t^{\ast}\right)  \geq T\left(  \left.  t^{\ast}\right\vert y_{0}%
,\tau\mu_{N-k},c_{U}\left(  k\right)  \right)  & \text{if }\Psi_{k}\left(
t^{\ast};\tau\right)  \leq c_{U}\left(  k\right)  \text{ for }t^{\ast}%
\in\left[  t_{1}^{\ast},t_{2}^{\ast}\right]  \text{ }\\
y\left(  t^{\ast}\right)  \leq T\left(  \left.  t^{\ast}\right\vert y_{0}%
,\tau\mu_{N-k},c_{L}\left(  k\right)  \right)  & \text{if }\Psi_{k}\left(
t^{\ast};\tau\right)  \geq c_{L}\left(  k\right)  \text{ for }t^{\ast}%
\in\left[  t_{1}^{\ast},t_{2}^{\ast}\right]
\end{array}
\right.
\]
The upper bound $\Psi_{k}\left(  t^{\ast};\tau\right)  \leq c_{U}\left(
k\right)  $ implies that $y\left(  t^{\ast}\right)  \geq T\left(  \left.
t^{\ast}\right\vert y_{0},\tau\mu_{N-k},c_{U}\left(  k\right)  \right)  $ and
$T\left(  \left.  t^{\ast}\right\vert y_{0},\tau\mu_{N-k},c_{U}\left(
k\right)  \right)  $ is real if the discriminant in (\ref{def_Xi}) is
positive,%
\[
\left(  1-\frac{1}{\tau\mu_{N-k}}\right)  ^{2}\geq\frac{4c}{\tau\mu_{N-k}}%
\geq\frac{4\Psi_{k}\left(  t^{\ast};\tau\right)  }{\tau\mu_{N-k}}=4\left(
\text{Var}\left[  S\left(  t^{\ast};\tau\right)  \right]  -\frac{E\left[
R_{k}\left(  t^{\ast};\tau\right)  \right]  }{N\mu_{N-k}}\right)
\]
If Var$\left[  S\left(  t^{\ast};\tau\right)  \right]  \geq\frac{E\left[
R_{k}\left(  t^{\ast};\tau\right)  \right]  }{N\mu_{N-k}}$ (i.e. $c_{L}\left(
k\right)  \geq0$), then the inequality is equivalent to the lower bound for
the effective infection rate
\begin{equation}
\tau\geq\frac{1}{\mu_{N-k}\left(  1-2\sqrt{\text{Var}\left[  S\left(  t^{\ast
};\tau\right)  \right]  -\frac{E\left[  R_{k}\left(  t^{\ast};\tau\right)
\right]  }{N\mu_{N-k}}}\right)  }\geq\frac{1}{\mu_{N-k}}
\label{tau_guaranteed_positive_prevalence}%
\end{equation}
that guarantees to operate in the endemic regime when $t^{\ast}$ is
sufficiently large. We observe that, the more connected components a graph on
$N$ nodes has, the larger $k$ and $\mu_{N-k}$ and, consequently, the lower
$\frac{1}{\mu_{N-k}}$. Physically, the larger $k$, the fewer nodes a connected
component has (on average $k/N$) and the larger the epidemic threshold of a
connected component should become, because $\tau_{c}>\tau_{c}^{\left(
1\right)  }=\frac{1}{\lambda_{1}}$ increases with decreasing $\lambda_{1}$ and
the spectral radius $\lambda_{1}\geq E\left[  D\right]  $, which, in dense
graphs, increases with $N$ on average. Hence, we expect that the epidemic
threshold in a graph with $k$ connected components increases and
(\ref{tau_guaranteed_positive_prevalence}) would imply that Var$\left[
S\left(  t^{\ast};\tau\right)  \right]  -\frac{E\left[  R_{k}\left(  t^{\ast
};\tau\right)  \right]  }{N\mu_{N-k}}\rightarrow\frac{1}{4}$ for sufficiently
large $t^{\ast}$.

In summary, the analysis generalizes the previous derivations in
\cite{PVM_bounds_SIS_prevalence} to graphs with $k$ connected components, as
e.g. in temporal networks. We can thus conclude that, for any graph $G$ with
$k$ connected components with a fixed topology in some non-zero time interval,
the prevalence $y\left(  t^{\ast}\right)  $ in that time interval can be
bounded by the curve $T\left(  \left.  t^{\ast}\right\vert y_{0},s,c\right)  $
in (\ref{Riccati_approximation_prevalence_k_connected_component_graph}) with
three parameters: (a) the initial condition $y_{0}$ or value of the prevalence
at the beginning of the time interval, (b) a Laplacian normalized rate
$s=\tau\mu_{N-k}$ and (c) a constant $c$. Implicitly, the computation of the
prevalence also assumes that the number $N$ of nodes in the graph $G$ is
known. The prevalence is only non-zero when the effective infection rate
$\tau$ exceeds the epidemic threshold $\tau_{c}$. Moreover, it is known that
$\tau>\tau_{c}>\tau_{c}^{\left(  1\right)  }$, where the NIMFA threshold
$\tau_{c}^{\left(  1\right)  }=\frac{1}{\lambda_{1}}$ and $\lambda_{1}$ is the
largest eigenvalue of the adjacency matrix of the graph $G$. Hence, the
adjacency normalized effective infection rate $x=\lambda_{1}\tau$ allows us to
compare epidemics in different graphs for sufficiently long time: when
$x\leq1$, the epidemic will die out, whereas $x>1+\varepsilon$ with
$\varepsilon$ a correction due to the mean-field approximation, the epidemic
will be persistent. However, the correction $\varepsilon$ is unknown. On the
other hand, (\ref{tau_guaranteed_positive_prevalence}) tells us that, when the
Laplacian normalized rate $s=\tau\mu_{N-k}\geq\xi$, where $\xi=\frac
{1}{1-2\sqrt{\text{Var}\left[  S\left(  t^{\ast};\tau\right)  \right]
-\frac{E\left[  R_{k}\left(  t^{\ast};\tau\right)  \right]  }{N\mu_{N-k}}}}%
>1$, we are surely in the endemic regime where the prevalence $y\left(
t^{\ast}\right)  >0$ (for a sufficiently large $t^{\ast}$). Unfortunately,
computing $\xi$ is difficult, so that determining which effective infection
rate $\tau$ leads to persistent infections, is complicated.

In conclusion, an accurate determination of the SIS epidemics threshold regime
will likely stay on the scientific agenda for future research.

\section{Governing equation of the heterogeneous SIS prevalence in graph with
$k$ disconnected components}

\label{sec_tanh_heterogeneousSIS_k_components}The expression
(\ref{quadratic_form_wTQw_inS_k_connected_component_graph}) is valid for the
weighted Laplacian $\widetilde{Q}$ with eigenvectors $\widetilde{x}%
_{1},\widetilde{x}_{2},\ldots,\widetilde{x}_{N}=u$ belonging to eigenvalues
$\widetilde{\mu}_{1}\geq\widetilde{\mu}_{2}\geq\ldots\geq\widetilde{\mu}%
_{N}=0$, respectively and with the scalar product $\widetilde{\zeta}_{k}%
=w^{T}\widetilde{x}_{k}$, since the kernel space of $\widetilde{Q}$ is the
same as that of the unweighted Laplacian $Q$. Invoking the definition of the
prevalence $y=E\left[  S\right]  $, $E\left[  S-S^{2}\right]  =y-E\left[
S^{2}\right]  $ and $E\left[  S^{2}\right]  =y^{2}+$ Var$\left[  S\right]  $,
(\ref{quadratic_form_wTQw_inS_k_connected_component_graph}) becomes%
\[
\frac{1}{N}E\left[  w^{T}\widetilde{Q}w\right]  =\widetilde{\mu}%
_{N-k}y-\widetilde{\mu}_{N-k}y^{2}-\widetilde{\mu}_{N-k}\left(  \text{Var}%
\left[  S\right]  -\frac{E\left[  \widetilde{R}_{k}\right]  }{N\widetilde{\mu
}_{N-k}}\right)
\]
Introduced into (\ref{dvgl_heterogeneous_prevalence}), which we write as,
\begin{align*}
\frac{dy\left(  t;\tau\right)  }{dt}  &  =-\frac{1}{N}E\left[  \left(
\delta_{av}u+\widetilde{\delta}-\delta_{av}u\right)  ^{T}w\left(  t\right)
\right]  +\frac{1}{N}E\left[  \left(  w\left(  t\right)  \right)
^{T}\widetilde{Q}w\left(  t\right)  \right] \\
&  =-\delta_{av}y-\frac{1}{N}E\left[  \left(  \widetilde{\delta}^{T}%
-\delta_{av}u^{T}\right)  w\left(  t\right)  \right]  +\widetilde{\mu}%
_{N-k}y-\widetilde{\mu}_{N-k}y^{2}-\widetilde{\mu}_{N-k}\left(  \text{Var}%
\left[  S\right]  -\frac{E\left[  \widetilde{R}_{k}\right]  }{N\widetilde{\mu
}_{N-k}}\right)
\end{align*}
where $\left(  \widetilde{\delta}^{T}-\delta_{av}u^{T}\right)  w\left(
t\right)  $ is now an additional correction due to heterogeneous,
node-depending curing rates. Normalizing the time $\widetilde{t^{\ast}}%
=\delta_{av}t$ with respect to average curing rate $\delta_{av}=\frac
{\widetilde{\delta}^{T}u}{N}$ and realizing that the eigenvalues of the
weighted Laplacian are function of the heterogeneous infection rates
$\beta_{ij}$, we have%
\[
\frac{dy\left(  \widetilde{t^{\ast}};\tau\right)  }{d\widetilde{t^{\ast}}%
}=\left(  \frac{\widetilde{\mu}_{N-k}}{\delta_{av}}-1\right)  y-\frac
{\widetilde{\mu}_{N-k}}{\delta_{av}}y^{2}-\frac{\widetilde{\mu}_{N-k}}%
{\delta_{av}}\left(  \text{Var}\left[  S\right]  -\frac{E\left[  \widetilde
{R}_{k}\right]  -E\left[  \left(  \widetilde{\delta}^{T}-\delta_{av}%
u^{T}\right)  w\left(  t\right)  \right]  }{N\widetilde{\mu}_{N-k}}\right)
\]
In summary, the spectral representation of the heterogeneous SIS prevalence
governing differential equation in a graph with $k$ connected components is%
\[
\frac{dy\left(  \widetilde{t^{\ast}};\tau\right)  }{d\widetilde{t^{\ast}}%
}=\left(  \frac{\widetilde{\mu}_{N-k}}{\delta_{av}}-1\right)  y-\frac
{\widetilde{\mu}_{N-k}}{\delta_{av}}y^{2}-\widetilde{\Psi}_{k}\left(  t^{\ast
};\tau\right)
\]
where the remainder is%
\[
\widetilde{\Psi}_{k}\left(  t^{\ast};\tau\right)  =\frac{\widetilde{\mu}%
_{N-k}}{\delta_{av}}\left(  \text{Var}\left[  S\right]  -\frac{E\left[
\widetilde{R}_{k}\right]  -E\left[  \left(  \widetilde{\delta}^{T}-\delta
_{av}u^{T}\right)  w\left(  t\right)  \right]  }{N\widetilde{\mu}_{N-k}%
}\right)
\]
Just as in the homogeneous case (Appendix
\ref{sec_tanh_homogenousSIS_k_components}), we may proceed by a bounding
procedure to find that the heterogeneous SIS prevalence also can be bounded by
a tanh-expression of the form
(\ref{Riccati_approximation_prevalence_k_connected_component_graph}), though
with different coefficients and an even more complicated $\widetilde{\Psi}%
_{k}\left(  t^{\ast};\tau\right)  $.

\section{Governing equation of the $\varepsilon-$SIS prevalence in graph with
$k$ disconnected components}

\label{sec_tanh_epsilonSIS_k_components}The differential equation for the
average fraction of infected nodes $y$ in the $\varepsilon-$SIS process is
\cite[p. 455]{PVM_PAComplexNetsCUP}
\begin{equation}
\frac{dy\left(  t^{\ast};\tau\right)  }{dt^{\ast}}=\varepsilon^{\ast}-\left(
1+\varepsilon^{\ast}\right)  y\left(  t^{\ast};\tau\right)  +\frac{\tau}%
{N}E\left[  w^{T}\left(  t^{\ast};\tau\right)  Qw\left(  t^{\ast};\tau\right)
\right]  \label{differential_equation_epsilonSISprevalence_Laplacian}%
\end{equation}
where $\varepsilon$ is the constant self-infection rate for each node
\cite{PVM_EpsilonSIS_PRE2012}. If $\tau=0$, the differential equation
(\ref{differential_equation_epsilonSISprevalence_Laplacian}) for the
$\varepsilon$-SIS prevalence shows that%
\[
\frac{dy\left(  t^{\ast};0,\varepsilon^{\ast}\right)  }{dt^{\ast}}%
=\varepsilon^{\ast}-\left(  1+\varepsilon^{\ast}\right)  y\left(  t^{\ast
};0,\varepsilon^{\ast}\right)
\]
with solution%
\[
y\left(  t^{\ast};0,\varepsilon^{\ast}\right)  =\frac{\varepsilon^{\ast}%
}{1+\varepsilon^{\ast}}+\left(  y_{0}-\frac{\varepsilon^{\ast}}{1+\varepsilon
^{\ast}}\right)  e^{-\left(  1+\varepsilon^{\ast}\right)  t}%
\]

As in previous Section \ref{sec_tanh_homogenousSIS_k_components}, after using
$y=E\left[  S\right]  $, $E\left[  S-S^{2}\right]  =y-E\left[  S^{2}\right]  $
and $E\left[  S^{2}\right]  =y^{2}+$ Var$\left[  S\right]  $ and
(\ref{quadratic_form_wTQw_inS_k_connected_component_graph}), the differential
equation (\ref{differential_equation_epsilonSISprevalence_Laplacian}) of the
$\varepsilon$-SIS prevalence becomes%
\[
\frac{dy\left(  t^{\ast};\tau,\varepsilon^{\ast}\right)  }{dt^{\ast}}=\left(
\tau\mu_{N-k}-\left(  1+\varepsilon^{\ast}\right)  \right)  y\left(  t^{\ast
};\tau,\varepsilon^{\ast}\right)  -\tau\mu_{N-k}y^{2}\left(  t^{\ast}%
;\tau,\varepsilon^{\ast}\right)  -\Psi_{k}\left(  t^{\ast};\tau,\varepsilon
^{\ast}\right)
\]
where%
\begin{equation}
\Psi_{k}\left(  t^{\ast};\tau,\varepsilon^{\ast}\right)  =\tau\mu_{N-k}\left(
\text{Var}\left[  S\left(  t^{\ast};\tau\right)  \right]  -\frac{E\left[
R_{k}\left(  t^{\ast};\tau\right)  \right]  }{N\mu_{N-k}}\right)
-\varepsilon^{\ast} \label{remainder_epsilon_SIS}%
\end{equation}
Again, by bounding $c_{L}\left(  k;\varepsilon^{\ast}\right)  \leq\Psi
_{k}\left(  t^{\ast};\tau,\varepsilon^{\ast}\right)  \leq c_{U}\left(
k;\varepsilon^{\ast}\right)  $, a variant of the tanh-formula
(\ref{Riccati_approximation_prevalence_k_connected_component_graph}) applies%
\begin{equation}
\widetilde{T}\left(  \left.  t^{\ast}\right\vert y_{0},s,c;\varepsilon^{\ast
}\right)  =\frac{1}{2}\left(  1-\frac{1+\varepsilon^{\ast}}{s}\right)
+\frac{\Upsilon}{2}\tanh\left(  \frac{s\Upsilon}{2}t^{\ast}+\text{arctanh}%
\left(  \frac{2y_{0}-\left(  1-\frac{1+\varepsilon^{\ast}}{s}\right)
}{\Upsilon}\right)  \right)
\label{Riccati_approximation_epsilonSISprevalence_k_connected_component_graph}%
\end{equation}
where $s=\tau\mu_{N-k}$ and
\[
\Upsilon=\sqrt{\left(  1-\frac{1+\varepsilon^{\ast}}{s}\right)  ^{2}-\frac
{4c}{s}}%
\]
which clearly reduces to
(\ref{Riccati_approximation_epsilonSISprevalence_k_connected_component_graph})
for $\varepsilon^{\ast}=0$.

\subsection{Extremal values of the $\varepsilon$-SIS prevalence}

\label{sec_extremal_values_epsilonSIS}When the prevalence attains an extremum
$y\left(  p;\tau,\varepsilon^{\ast}\right)  $ at time $t^{\ast}=p$, obeying
$\left.  \frac{dy\left(  t^{\ast};\tau,\varepsilon^{\ast}\right)  }{dt^{\ast}%
}\right\vert _{t^{\ast}=p}=0$, then%
\[
\tau\mu_{N-k}y^{2}\left(  p;\tau,\varepsilon^{\ast}\right)  -\left(  \tau
\mu_{N-k}-\left(  1+\varepsilon^{\ast}\right)  \right)  y\left(
p;\tau,\varepsilon^{\ast}\right)  +\Psi_{k}\left(  p;\tau,\varepsilon^{\ast
}\right)  =0
\]
There are only real solutions for the \textquotedblleft
time-extremal\textquotedblright\ prevalence,
\[
y_{\pm}\left(  p;\tau,\varepsilon^{\ast}\right)  =\frac{\left(  \tau\mu
_{N-k}-\left(  1+\varepsilon^{\ast}\right)  \right)  \pm\sqrt{\left(  \tau
\mu_{N-k}-\left(  1+\varepsilon^{\ast}\right)  \right)  ^{2}-4\tau\mu
_{N-k}\Psi_{k}\left(  p;\tau,\varepsilon^{\ast}\right)  }}{2\tau\mu_{N-k}}%
\]
and%
\begin{equation}
y_{\pm}\left(  p;\tau,\varepsilon^{\ast}\right)  =\left(  1-\frac
{1+\varepsilon^{\ast}}{\tau\mu_{N-k}}\right)  \frac{1}{2}\left\{  1\pm
\sqrt{1-\frac{\frac{\Psi_{k}\left(  p;\tau,\varepsilon^{\ast}\right)  }%
{\tau\mu_{N-k}}}{1-\frac{1+\varepsilon^{\ast}}{\tau\mu_{N-k}}}}\right\}
\label{time_extremal_epsilon_SIS_prevalence}%
\end{equation}
provided the discriminant $\left(  \tau\mu_{N-k}-\left(  1+\varepsilon^{\ast
}\right)  \right)  ^{2}-4\tau\mu_{N-k}\Psi_{k}\left(  p;\tau,\varepsilon
^{\ast}\right)  \geq0$, which is equivalent to%
\[
\frac{1}{4}\left(  1-\frac{1+\varepsilon^{\ast}}{\tau\mu_{N-k}}\right)
^{2}\geq\frac{\Psi_{k}\left(  p;\tau,\varepsilon^{\ast}\right)  }{\tau
\mu_{N-k}}=\text{Var}\left[  S\left(  p;\tau\right)  \right]  -\frac{E\left[
R_{k}\left(  p;\tau\right)  \right]  }{N\mu_{N-k}}-\frac{\varepsilon^{\ast}%
}{\tau\mu_{N-k}}%
\]
This inequality leads to a lower bound for the effective infection rate,%
\[
\tau\left(  \varepsilon^{\ast}\right)  \geq\frac{1}{\mu_{N-k}}\frac
{1+\varepsilon^{\ast}}{1-2\sqrt{\text{Var}\left[  S\left(  p;\tau\right)
\right]  -\frac{E\left[  R_{k}\left(  p;\tau\right)  \right]  }{N\mu_{N-k}%
}-\frac{\varepsilon^{\ast}}{\tau\mu_{N-k}}}}>\tau\left(  0\right)
\]
For small self-infection rates $\varepsilon^{\ast}$, we can demonstrate the
last inequality, which implies that the effective infection rate to guarantee
an endemic regime lies higher for the $\varepsilon$-SIS model than for the
classical ($\varepsilon=0$) SIS-model, which is consistent with the analysis
in \cite[p. 457-458]{PVM_PAComplexNetsCUP}. Although surprising at first
glance, we need to realize that the steady state in the $\varepsilon=0$
SIS-model is the overall healthy state, to which the $\varepsilon$-SIS must
converge if $\varepsilon\rightarrow0$, irrespective of the effective infection
rate $\tau$. The existence of an absorbing state implies in a finite graph
that the epidemics eventually dies out (i.e. the dynamic process will surely
hit the absorbing state in the $2^{N}$ large Markov state graph). This
peculiar limit $\varepsilon\rightarrow0$ is further illustrated in \cite[Fig.
5 and 6]{PVM_EpsilonSIS_PRE2012}.

\subsection{The absorbing state}

\label{sec_absorbing_state_epsilonSIS}The \textquotedblleft
time-extremal\textquotedblright\ prevalence $y_{\pm}\left(  p;\tau
,\varepsilon^{\ast}\right)  $ at time $t^{\ast}=p$ in
(\ref{time_extremal_epsilon_SIS_prevalence}) can only be zero if (a) the
negative sign applies and (b) $\Psi_{k}\left(  p;\tau,\varepsilon^{\ast
}\right)  =0$, in which case $y_{-}\left(  p;\tau,\varepsilon^{\ast}\right)
=0$ for all effective infection rates $\tau$. However, if $y_{\pm}\left(
p;\tau,\varepsilon^{\ast}\right)  =0$, then $S\left(  p\right)  =0$, which
implies that Bernoulli vector $w\left(  p\right)  =0$. The definition
(\ref{remainder_epsilon_SIS}) of the remainder $\Psi_{k}\left(  p;\tau
,\varepsilon^{\ast}\right)  $ and the specific expression
(\ref{general_R_k_correction}) for the spectral correction $R_{k}$ illustrate
that $\Psi_{k}\left(  p;\tau,\varepsilon^{\ast}\right)  =-\varepsilon^{\ast}$
if $S\left(  p\right)  =0$. Consequently, $y_{-}\left(  p;\tau,\varepsilon
^{\ast}\right)  =0$ and $\Psi_{k}\left(  p;\tau,\varepsilon^{\ast}\right)  =0$
can only be satisfied if $\varepsilon^{\ast}=0$, thus only in the classical
SIS process. This singular condition, $y_{-}\left(  p;\tau,\varepsilon^{\ast
}\right)  =0$ and $\Psi_{k}\left(  p;\tau,\varepsilon^{\ast}\right)  =0$,
which holds irrespective of the effective infection rate $\tau$, corresponds
to the absorbing state which is attained at time $p$. If $\varepsilon^{\ast
}>0$, there cannot be an absorbing state and the negative sign solution
$y_{-}\left(  p;\tau,\varepsilon^{\ast}\right)  $ in
(\ref{time_extremal_epsilon_SIS_prevalence}), which is decreasing in $\tau$,
does not exist. Moreover, Markov theory \cite{PVM_PAComplexNetsCUP} states the
$\varepsilon$-SIS Markovian chain possesses a unique steady-state, which
corresponds to $y_{+}\left(  p;\tau,\varepsilon^{\ast}\right)  $ in
(\ref{time_extremal_epsilon_SIS_prevalence}). When bounding $\Psi_{k}\left(
t^{\ast};\tau,\varepsilon^{\ast}\right)  \geq c_{L}$ so as to prevent that
$\Psi_{k}\left(  p;\tau,\varepsilon^{\ast}\right)  \rightarrow0$, then the
limit $\varepsilon^{\ast}\rightarrow0$ in
(\ref{time_extremal_epsilon_SIS_prevalence}) will correspond to the metastable
state of the SIS process,%
\[
y_{\pm}\left(  p;\tau,0\right)  =\frac{1}{2}\left(  1-\frac{1}{\tau\mu_{N-k}%
}\right)  \left\{  1+\sqrt{1-\frac{\frac{c_{L}}{\tau\mu_{N-k}}}{1-\frac
{1}{\tau\mu_{N-k}}}}\right\}
\]
which is precisely equal to the tanh-formula's steady-state
(\ref{tanh_steady_state}).

\end{document}